\documentclass[]{aastex631}

\begin{document}

\title{B2 1308+326: a changing-look blazar or not?}
\correspondingauthor{Ashwani Pandey}
\email{ashwanitapan@gmail.com}

\author{Ashwani Pandey}\thanks{PIFI visiting scientist}
\affiliation{Key Laboratory for Particle Astrophysics, Institute of High Energy Physics, Chinese Academy of Sciences, 19B Yuquan Road, Beijing 100049, P. R. China}
\affiliation{Department of Physics and Astronomy, University of Utah, Salt Lake City, UT 84112, USA}
\affiliation{Center for Theoretical Physics, Polish Academy of Sciences, Al.Lotnikov 32/46, PL-02-668 Warsaw, Poland}
\author{Chen Hu}
\affiliation{Key Laboratory for Particle Astrophysics, Institute of High Energy Physics, Chinese Academy of Sciences, 19B Yuquan Road, Beijing 100049, P. R. China}

\author{Jian-Min Wang}
\affiliation{Key Laboratory for Particle Astrophysics, Institute of High Energy Physics, Chinese Academy of Sciences, 19B Yuquan Road, Beijing 100049, P. R. China}
\affil{School of Astronomy and Space Science, University of Chinese Academy of
Sciences, Beijing 100049, China}
\affil{National Astronomical Observatories of China, The Chinese
Academy of Sciences, 20A Datun Road, Beijing 100020, China}

\author{Bo\.zena Czerny}
\affiliation{Center for Theoretical Physics, Polish Academy of Sciences, Al.Lotnikov 32/46, PL-02-668 Warsaw, Poland}

\author{Yong-Jie Chen}
\affiliation{Key Laboratory for Particle Astrophysics, Institute of High Energy Physics, Chinese Academy of Sciences, 19B Yuquan Road, Beijing 100049, P. R. China}
\affil{Dongguan Neutron Science Center, 1 Zhongziyuan Road, Dongguan 523808, China}

\author{Yu-Yang Songsheng}
\affiliation{Key Laboratory for Particle Astrophysics, Institute of High Energy Physics, Chinese Academy of Sciences, 19B Yuquan Road, Beijing 100049, P. R. China}

\author{Yi-Lin Wang}
\affiliation{Key Laboratory for Particle Astrophysics, Institute of High Energy Physics, Chinese Academy of Sciences, 19B Yuquan Road, Beijing 100049, P. R. China}
\affiliation{School of Physical Science, University of Chinese Academy of Sciences, 19A Yuquan Road, Beijing 100049, People's Republic of China}

\author{Hao Zhang}
\affiliation{Key Laboratory for Particle Astrophysics, Institute of High Energy Physics, Chinese Academy of Sciences, 19B Yuquan Road, Beijing 100049, P. R. China}
\affiliation{School of Physical Science, University of Chinese Academy of Sciences, 19A Yuquan Road, Beijing 100049, People's Republic of China}

\author{Jes\'us Aceituno}
\affil{Centro Astronomico Hispano Alem\'an, Sierra de los filabres sn, E-04550 Gergal, Almer\'ia, Spain}
\affil{Instituto de Astrof\'isica de Andaluc\'ia (CSIC), Glorieta de la
astronom\'ia sn, E-18008 Granada, Spain}

\begin{abstract}
In our previous study, we identified a shift in the synchrotron peak frequency of the blazar B2 1308$+$326 from 10$^{12.9}$ Hz to 10$^{14.8}$ Hz during a flare, suggesting it could be a changing-look blazar (CLB). In this work, we investigate the CL behaviour of B2 1308+326 by analysing a newly acquired optical spectrum and comparing it with an archival spectrum. We find that between the two epochs, the continuum flux increased by a factor of $\sim$4.4, while the Mg II emission line flux decreased by a factor of  1.4$\pm$0.2. Additionally, the equivalent width of the Mg II line reduced from $\sim 20$ \AA \ to $\sim 3$ \AA, indicating an apparent shift from a flat-spectrum radio quasar (FSRQ) class to a BL Lacertae (BL Lac) class. Despite this apparent change, the ratio of accretion disk luminosity to Eddington luminosity remains $>$ 10$^{-2}$ during both epochs, indicating efficient accretion persists in B2 1308$+$326. 
The measured black hole mass remains consistent with an average $\log M_{\rm BH} = 8.44$ M$_{\odot}$. 
Our findings suggest that B2 1308$+$326 is not a genuine CLB, but rather an intrinsic FSRQ that emerges as a BL Lac during high-flux states due to enhanced non-thermal emission.

\end{abstract}


\section{Introduction} \label{sec:intro}
Broad emission lines (BELs) and the underlying thermal continuum from the accretion disk (AD) are the two key components of the UV/optical spectra of active galactic nuclei (AGN). These emission lines are formed due to the photoionization of broad line region (BLR) clouds by the incident continuum and are Doppler broadened. Based on the existence or lack of BELs in the spectra, AGN are generally divided into Type 1 and Type 2 AGN \citep[e.g.][]{Antonucci1993}. However, over the last ten years, an increasing number of sources have been discovered to exhibit a transition between Type 1 and Type 2 AGN and are referred to as changing-look AGN (CLAGN). 
The change in their spectral states is mostly due to the change in the underlying incident continuum from the AD but it could also be due to the change in the line-of-sight column density (see \cite{Ricci2023} for an excellent review).

Jetted-AGN\footnote{AGN having strong relativistic jets \citep{Padovani2017}} have an additional, highly variable, non-thermal continuum component due to the synchrotron emission by relativistic electrons within the jet that can affect their UV/optical spectra. 
Blazars are jetted-AGN with jets orientated closer to the observer \citep{Urry1995}. They are usually categorised into flat spectrum radio quasars (FSRQs) and BL Lacertae objects (BL Lacs) based on their UV/optical spectra; FSRQs have BELs (EW $>$ 5 \AA) in their spectra while the spectra of BL Lacs are mostly featureless \citep{Stocke1991}. A physical distinction between these two categories of blazars is proposed by \cite{Ghisellini2011} based on the ratio of disc luminosity ($L_{AD}$) to the Eddington luminosity ($L_{Edd}$). They suggested that BL Lacs have radiatively ineffective accretion ($L_{AD}/L_{Edd} < 10^{-2}$) and hence no emission lines, while in FSRQs, accretion is radiatively efficient, producing BELs in their spectra.  
The peak, $\nu_p$, of the low energy (synchrotron) component of the spectral energy distributions (SEDs) of blazar is another factor used to classify them. Based on  $\nu_p$, blazars are categorised as; low-synchrotron peaked (LSPs; $\nu_p \leq 10^{14}$ Hz), intermediate-synchrotron peaked (ISPs; 10$^{14} < \nu_p < 10^{15}$ Hz), and high-synchrotron peaked (HSPs; $\nu_p \geq 10^{15}$Hz) blazars \citep{Abdo2010}.

Similar to CLAGN, a number of blazars that transitioned from one blazar class to another over various observation epochs have been identified using changes in the emission line widths and also using shifts in the synchrotron peak frequency \citep[e.g.][]{Ghisellini2011,Ghisellini2013,Ruan2014,Mishra2021,Pandey2024,Paiano2024}.
 However, it is still unclear whether these sources are genuine changing look blazars (CLBs) or whether the transition is apparent because of increased jet continuum emission. During periods of high activity, the enhanced non-thermal jet continuum could swamp out the emission lines in the FSRQ spectrum, giving the impression that the source is a BL Lac  \citep[e.g][]{Ruan2014}. Similarly, a change in the Doppler factor, most likely due to a change in the viewing angle, could also shift the synchrotron peak \citep[e.g.][]{Pandey2024}.

B2 1308$+$326 (OP 313) is a high redshift blazar located at z=0.9980$\pm$0.0005 \citep{2010MNRAS.405.2302H}.
In our previous work, \cite{Pandey2024} (Hereafter, Paper 1), we examined the multiwavelength emission of B2 1308$+$326. Using broadband SED modelling, we found that its synchrotron peak, $\nu_p$, shifted from 10$^{12.9}$ Hz (in the low state) to 10$^{14.8}$ Hz (in the high state), indicating a change from LSP to ISP class. This encourages us to investigate its spectral characteristics to validate the transitional behaviour. In this work, we examined any variation in the emission line characteristics and underlying continuum flux of B2 1308$+$326 using two optical spectra (in the observed frame) obtained during different epochs.

The outline of this paper is as follows: Section \ref{sec:obs} describes the observations and data reduction process. Results are given in Section \ref{sec:res}. We discussed our findings in Section \ref{sect:discussion} and a summary of the work is presented in Section \ref{sec:summary}.
\begin{figure*}
    \centering
    \includegraphics[width=16cm, height=10cm]{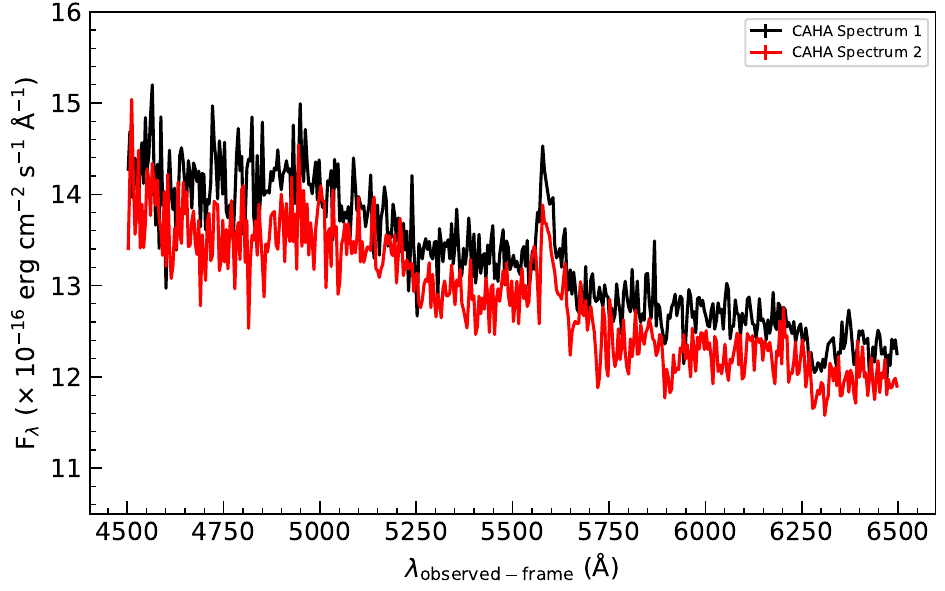}
    \caption{\label{fig:compare_CAHA}A comparison of two CAHA spectra, each with 
 an exposure time of $\sim$1200 s, taken on MJD 60479. We limited the spectra to the wavelength range of 4500–6500 \AA, as it contains high-quality data.}
\end{figure*}
\section{Observations and data reduction} \label{sec:obs}
\begin{figure*}
    \centering
    \includegraphics[width=16cm, height=10cm]{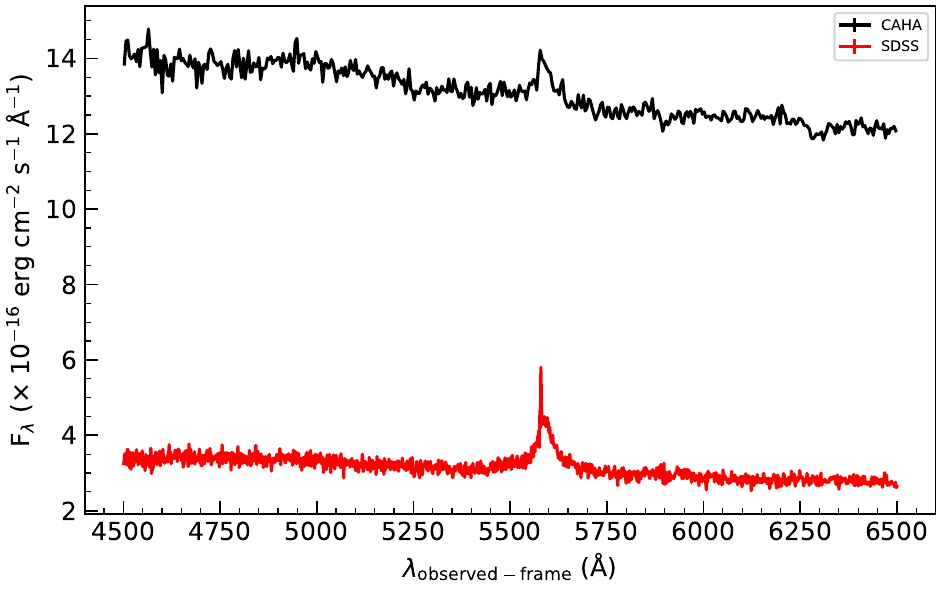}
    \caption{\label{fig:compare}A comparison of observed SDSS and CAHA spectra over the wavelength range 4500-6500 \AA}.
\end{figure*}

We examined two spectra of B2 1308$+$326; one archival Sloan Digital Sky Survey (SDSS) spectrum\footnote{taken from \url{https://cas.sdss.org/dr18/VisualTools/quickobj}} and one new observation with the Centro Astron\'{o}mico Hispano-Alem\'{a}n (CAHA) telescope. The archival SDSS spectrum was taken on 2006 March 25 (MJD 53819) and corresponds to plate ID 2029 and fiber number 602. The new observation was performed on 2024 June 18 (MJD 60479) at the Calar Alto Observatory. Two spectra of 1200 s each were taken under good weather
conditions using the Calar Alto Faint Object Spectrograph (CAFOS) mounted on the  2.2 m CAHA telescope. The Grism G-200 and a long slit set to a projected width of
3$\farcs$0 were employed. The night was a photometric night with a recorded seeing of 1.2$^{\prime\prime}$ when the spectra were acquired. However, we didn't take any images during these observations. We obtained two spectrophotometric standards one (BD+26 2606) at the beginning and one (BD+28d4211) at the end of the same night. The sensitivity curves from these standards show a minor difference of approximately 0.1 mag. Additionally, four other objects observed that night for our reverberation mapping campaign were calibrated using the same spectrophotometric standards as this target. A comparison of their fluxes with those calibrated using a simultaneously observed comparison star (for our reverberation mapping) showed differences of 15\%, 10\%, 6\%, and 2\% for the four objects, respectively \citep[e.g.][]{Hu2021}. Considering these comparisons, it is reasonable to say that our absolute flux calibration for this target has an accuracy of about 10\%. The data reduction was performed by the Image Reduction and Analysis Facility (IRAF) \footnote{IRAF is distributed by the National Optical Astronomy Observatories,
which are operated by the Association of Universities for Research in Astronomy, Inc., under a cooperative agreement with the National Science Foundation.} following the standard procedures. The reduced spectra cover the wavelength range of $\sim$4000--8500 \AA\ with a dispersion of 4.47 \AA\ pixel$^{-1}$. The instrumental broadening is $\sim$1000 $\rm km~s^{-1}$, estimated by
\citet{hu20} for the same set of Grism and slit width. These spectra are nearly identical, except for a difference in the shape in the extreme blue region around 4000 \AA. Additionally, telluric absorption lines appear at longer wavelengths, specifically beyond 6500 \AA. The wavelength range between 4500-6500 \AA \ shows good-quality data and is used for further analysis. We plotted the two CAHA spectra taken on MJD 60479 in Figure \ref{fig:compare_CAHA}. Each CAHA spectrum has a signal-to-noise (S/N) ratio of approximately 60 at the continuum around the Mg II line. By combining the two spectra, we achieved an improved S/N ratio of $\sim$85. Figure \ref{fig:compare} illustrates a direct comparison between the observed SDSS and CAHA spectra spanning the wavelength range of 4500–6500 Å, which includes good-quality data and is used for further investigations.

 
\begin{figure*}
    \centering
    \includegraphics[width=8cm, height=8.5cm, angle =-90]{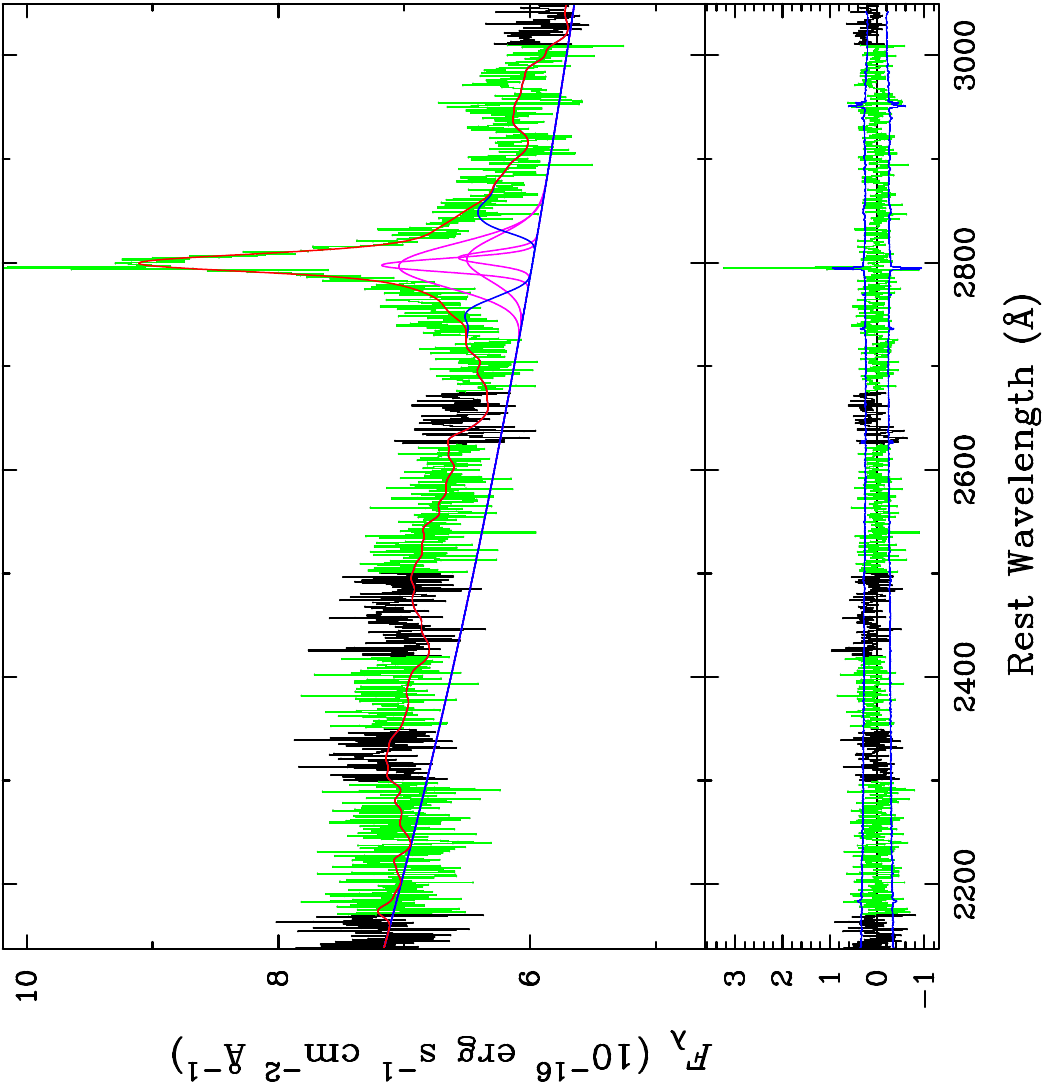}\hfill \includegraphics[width=8cm, height=8.5cm, angle =-90]{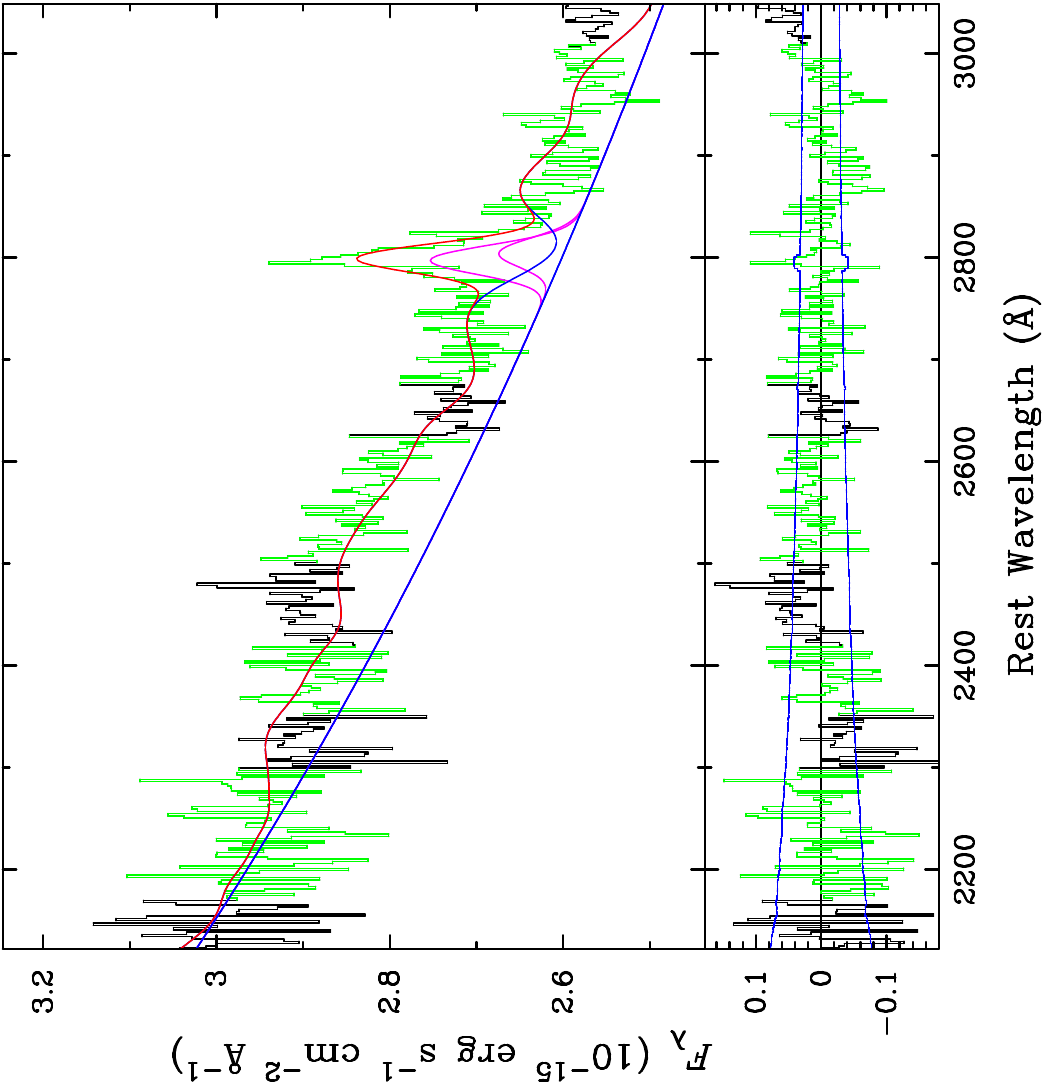}
    \caption{\label{fig:fitted_spectra}A fit to the spectra of B2 1308$+$326; Left panel: SDSS spectrum. Right panel: CAHA spectrum.}
\end{figure*}

\section{Results} \label{sec:res}
To examine the spectral characteristics of B2 1308$+$326, we first corrected both SDSS and CAHA spectra for Galactic extinction\footnote{A value of 0.039 for the $V$-band extinction magnitude from the NASA/IPAC Extragalactic Database \citep{schlafly11} was used.}. We then shift the observed spectra to the rest frame. The rest-frame spectra of B2 1308$+$326 have Mg II as the most prominent emission line. Finally, we performed the spectral fitting to both spectra to compute their different spectral components using the algorithm described in Appendix A of \citet{hu12}. Our model is similar to that used in \citet{hu08} and includes the following components: (1) the AGN continuum (a single power law of the form $F_{\lambda} \propto \lambda^{\alpha}$); (2) the Fe {\sc ii} pseudo continuum (modelled from the \citealt{vestergaard01} template); (3) the Mg II $\lambda\lambda$2796,2803 doublet (multi-Gaussian).

For the SDSS spectrum, each of the Mg II doublets is modelled using a pair of Gaussian functions: one representing the narrow component and the other capturing the broad component of the doublet \citep[e.g.][]{hu08,Shen2011}.
However, for the CAHA spectrum, a single Gaussian is enough for each doublet due to its relatively low spectral resolution. The doublets are forced to have the same widths and shifts, while the intensity ratio between them is fixed to the value of 2 \citep{baldwin96}. The FWHM of the Mg II line is calculated from the best-fit model of one of the Mg II doublets.  We corrected the observed FWHM of the Mg II line for the instrumental broadening using the following equation
\begin{equation}
    FWHM_{corrected} = \sqrt{FWHM^2_{observed}  - FWHM^2_{instrumental}} 
\end{equation}
The values of instrumental broadening for the SDSS and CAHA spectra are $\sim$167 and $\sim$1000 km/s, respectively.
The results of spectral fitting to the SDSS and CAHA spectra are presented in the left and right panels of Figure \ref{fig:fitted_spectra}, respectively. The corresponding fitting parameters are listed in Table \ref{tab:fit_par}. 

We also estimated the Mg II line flux using a simple integration method for the CAHA spectra, shown in the appendix \ref{app:A}. The resulting integrated flux is (7.3$\pm$0.4) $\times$ 10$^{-15}$ erg cm$^{-2}$ s$^{-1}$, which is slightly lower than the value obtained from the multiple Gaussian fitting. This difference could be due to overestimating the continuum level, as the Fe II emission was not subtracted which significantly influences the observed flux around the Mg II line \citep[e.g.][]{vestergaard01,baldwin2004,Pandey2024b}. Therefore, we proceed with the results of our fitting method.

We can estimate the virial mass of a black hole using the FWHM of the Mg II line and the continuum luminosity as \citep[e.g.][]{Mclure2002,Shaw2012}
\begin{equation}\label{eq:bh_mass}
   \log \left(\frac{M_{\rm BH}}{M_{\odot}} \right) = a + b \log \left(\frac{\lambda L_{\lambda}}{10^{44} {\rm erg } \  {\rm s}^{-1}} \right) + 2 \log \left( \frac{FWHM_{\rm Mg II}}{{\rm km} \ {\rm s}^{-1}}  \right) 
\end{equation}
where the coefficients a and b are calibrated using the reverberation mapping data.

Since in blazars (or jetted-AGN, in general), the UV/optical emission can be contaminated by the non-thermal jet emission \citep[e.g.][]{Wu2004,Liu2006}, in Equation \ref{eq:bh_mass} we used $L_{\rm Mg II}$ instead of $\lambda L_{\lambda}$ and adopted the corresponding coefficients ($a = 1.70,b = 0.63$) from \cite{Shaw2012}, to obtain M$_{\rm BH}$. From M$_{\rm BH}$, we got the Eddington luminosity as $L_{Edd} = $1.26 $\times$ 10$^{38}$ M$_{\rm BH}$/M$_{\odot}$ erg s$^{-1}$.  

The luminosity of the accretion disk, $L_{\rm AD}$, can be obtained from $L_{\rm Mg II}$, using the following scaling relationship \citep[e.g.][]{Zamaninasab2014}.
\begin{equation}
    \log L_{\rm AD} = (16.76 \pm 0.26) + (0.68 \pm 0.01) \log L_{\rm Mg II}
\end{equation}
We then computed the ratio $\lambda$ = $L_{\rm AD}/L_{Edd}$. 
The measured values of M$_{\rm BH}$, $L_{\rm AD}$ and $\lambda$ for the two spectra are given in Table \ref{tab:fit_par2}. 
Considering the significant uncertainties in the measurement of FWHM of the Mg II line, the values of M$_{\rm BH}$ are consistent for the two spectra. The values of $L_{\rm AD}$ are also marginally different for the two spectra. Both $\lambda$ values are larger than 10$^{-2}$, the threshold limit for a blazar to be categorised as an FSRQ.
\begin{table*}
   \centering 
    \caption{\label{tab:fit_par}Results of spectral fitting. The continuum fluxes are in the units of 10$^{-16}$ erg cm$^{-2}$ s$^{-1}$ \AA$^{-1}$, the Mg II line fluxes are in the units of 10$^{-16}$ erg cm$^{-2}$ s$^{-1}$, and Fe II line fluxes are in units of 10$^{-14}$ erg cm$^{-2}$ s$^{-1}$ .  }  
    \begin{tabular}{|c|c|c|c|c|c|c|c|} \hline
Spectra & Date of observation  & F$_{\rm 3000}$  & PL index & F$_{\rm Mg II}$  & EW$_{\rm Mg II}$ (\AA)  & FWHM (km s$^{-1}$) & F$_{\rm Fe II}$  \\ \hline
SDSS & 25 March 2006   & 5.71$\pm$0.08   & -0.67$\pm$0.02  &  112.93$\pm$8.28  &  19.78$\pm$1.48   &  2499.43$\pm$120.20  & 1.34$\pm$0.11 \\ \hline
CAHA & 18 June 2024   & 25.01$\pm$0.22   &  -0.54$\pm$0.01  & 79.99$\pm$6.31   & 3.20$\pm$0.25    &  3422.16$\pm$250.97  & 2.72$\pm$0.38 \\
\hline
\end{tabular}
\end{table*}
\begin{table*}
   \centering 
    \caption{\label{tab:fit_par2}The measured values of the luminosity of Mg II line ($L_{\rm Mg II}$), black hole mass ($M_{\rm BH}$), AD luminosity ($L_{\rm AD}$), and the ratio $L_{\rm AD}/L_{\rm Edd}$ for the two spectra.  }  
    \begin{tabular}{|c|c|c|c|c|} \hline
Spectra & $\log L_{\rm Mg II}$ (erg s$^{-1}$)  & $\log M_{\rm BH}$ (M$_{\odot}$) & $\log L_{\rm AD}$ (erg s$^{-1}$) & $\lambda$ = $L_{\rm AD}/L_{Edd}$ \\ \hline
SDSS    & 43.77$\pm$0.07  &  8.35$\pm$0.11 &  46.53$\pm$0.26  & 1.18$\pm$0.76 \\ \hline
CAHA    & 43.62$\pm$0.08  &  8.53$\pm$0.15 &  46.42$\pm$0.26  & 0.62$\pm$0.43 \\
\hline
\end{tabular}
\end{table*}
\section{Discussion}\label{sect:discussion}
A transition from one class to another has been observed in several blazars by observing changes in either the EW of broad emission lines or in the synchrotron peak frequency. Blazar B2 1308$+$326 underwent a transition from FSRQ to BL Lac category, which we identified by measuring an order of two shifts in its synchrotron peak frequency in our previous work (Paper I). A similar change in its synchrotron peak frequency was also noticed by \cite{Watson2000}. In this study, we investigated the transitional behaviour of B2 1308$+$326 by analysing two of its spectra, obtained with SDSS on MJD 53819 and with CAHA on MJD 60479. The rest-frame UV spectra of B2 1308$+$326 are characterised by the underlying PL continuum, broad Mg II line and Fe II emission.  We found that the value of continuum flux at 3000 \AA \ is increased from (5.71$\pm$0.08) $\times$ 10$^{-16}$ erg cm$^{-2}$ s$^{-1}$ \AA$^{-1}$ on MJD 53819  to (25.01$\pm$0.22) $\times$ 10$^{-16}$ erg cm$^{-2}$ s$^{-1}$ \AA$^{-1}$ on MJD 60479, accompanied by a change in the PL index from -0.67 to -0.54. However, the Mg II line flux decreased from (112.93$\pm$8.28) $\times$ 10$^{-16}$ erg cm$^{-2}$ s$^{-1}$ to (79.99$\pm$6.31) $\times$ 10$^{-16}$ erg cm$^{-2}$ s$^{-1}$, by a factor of 1.4$\pm$0.2, considering uncertainty ($\sim$10\%) in the photometric calibration. Due to this enhancement in the continuum flux by a factor of $\sim$4.4 and a decrease in Mg II line flux by a factor of $\sim$1.4, the EW$_{\rm Mg II}$ is decreased by a factor of $\sim$6 from  (19.78$\pm$1.48) to (3.20$\pm$0.25) \AA. Such a change in the EW$_{\rm Mg II}$ may indicate that the source shifted from FSRQ (EW $\geq$ 5 \AA) to BL Lac class (EW $<$ 5 \AA). Different values of EW$_{\rm Mg II}$ were reported, separately, for B2 1308$+$326 in the literature. \cite{Miller1978} observed EW$_{\rm Mg II} < 5$ \AA, while no emission lines were detected by \cite{Wills1979}. The larger values of EW$_{\rm Mg II} \sim 18.6$ \AA \ and   $\sim 15$ \AA \ were reported by \cite{Stickel1993} and \cite{Watson2000}, respectively, during the low luminosity states of the source. In this work, we also noticed that when the source was in a low state (MJD 53819), the EW$_{\rm Mg II} \geq 5$ \AA \, which is reduced to EW$_{\rm Mg II} < 5$ \AA \, when the source continuum increased. Thus, changes in the emission line properties of B2 1308$+$326 are predominantly due to the variations in its non-thermal jet continuum which consequently lead to an apparent change in its class. 

To further investigate the intrinsic nature of B2 1308$+$326, we computed the ratio  $\lambda$ = $L_{\rm AD}/L_{Edd}$.  A value of $\lambda \geq$ 10$^{-2}$ indicates that the AD is radiatively efficient to produce strong emission lines in the spectra and thereby represents an FSRQ source \citep{Ghisellini2011}. However, $\lambda <$ 10$^{-2}$ indicates radiatively inefficient accretion, meaning that there are weak or no emission lines, which points to a BL Lac source. We observed that $\lambda >$ 10$^{-2}$ for both the spectra of B2 1308$+$326 considered in this work. Thus, B2 1308$+$326 is intrinsically an FSRQ source that shows an apparent change from one class to another due to the fluctuations in its non-thermal jet continuum thereby pretending to be a CLB. 

We used the luminosity and FWHM of the Mg II line to calculate the $M_{\rm BH}$ of B2 1308$+$326 under the assumption of a virialized BLR. For the SDSS and CAHA spectra, the logarithmic values of $M_{\rm BH}$ are 8.35$\pm$0.11 M$_{\odot}$ and 8.53$\pm$0.15 M$_{\odot}$, respectively, which are comparable within the uncertainties. A slight increase in the velocity (FWHM) of the Mg II line could be due to the decrease in the Mg II line flux which consequently decreases the radius of the Mg II line forming region ($R \propto L^{1/2}$; \citep{2020ApJ...896..146Z}).

Additionally, we observed that, in contrast to the decline in the Mg II line flux, there is a hint of a slight increase in the Fe II flux with increasing continuum flux level, going from (1.34$\pm$0.11) $\times$ 10$^{-14}$ to (2.72$\pm$0.38) $\times$ 10$^{-14}$ erg cm$^{-2}$ s$^{-1}$, by a factor of 2$\pm$0.3. As the Fe II flux measurements are highly sensitive to the assumed continuum level, particularly in spectra with moderate S/N and spectral resolution, higher-quality spectra could help to confirm this opposing behaviour in Mg II and Fe II flux levels.
A similar trend was also observed in the CLB B2 1420$+$32 by \cite{2021A&A...647A.163M} and \cite{Mishra2021}. This is an unusual behaviour that might point to two distinct emitting regions for the Mg II and Fe II lines. It also suggests that the Fe II emitting gas may align close to the observer's line of sight and thereby interact with the jet emission. Another possible explanation for this unusual trend is that the dusty clouds may undergo sublimation with rising continuum flux, releasing a substantial amount of iron (Fe) into the surrounding environment \cite{Mishra2021}.

\section{Summary}\label{sec:summary}
In paper I, we noticed that the synchrotron peak frequency of B2 1308$+$326 has shifted from 10$^{12.9}$ Hz to 10$^{14.8}$ Hz during a flare, providing a hint that it could be a CLB. In this work, we 
observed a new spectrum of B2 1308$+$326 with the CAHA telescope and compared it with an archival SDSS spectrum to investigate its changing look behaviour.  The key findings of this work are as follows.
\begin{itemize}
    \item During the two epochs, the continuum flux (which is possibly a combination of thermal AD flux and non-thermal jet continuum) has increased by a factor of $\sim$4.4, while the Mg II line flux has decreased by a factor of $\sim$1.4. 
    \item The EW of the Mg II line decreased from $\sim 20$ \AA \ to $\sim 3$ \AA \ indicating an apparent change from FSRQ to BL Lac class. 
    \item However, the value of the ratio $\lambda$ = $L_{\rm AD}/L_{Edd}$ remains $ > 10^{-2}$ for the two spectra representing an efficient accretion in the source. Thus, the blazar B2 1308$+$326 is intrinsically an  FSRQ and the apparent change from FSRQ to BL Lac class is due to the increased non-thermal jet continuum which diluted the emission lines.
    \item Considering measurement uncertainties, the black hole mass of B2 1308$+$326 is consistent for the two spectra with an average value of $\log M_{\rm BH} = 8.44$ M$_{\odot}$.
    \item The Fe II flux has marginally increased by a factor of $\sim$2 with an increase in the continuum flux.
\end{itemize}
We propose that the blazar B2 1308$+$326 should not be regarded as a true CLB, but rather as an FSRQ that masquerades as a BL Lac source during high flux states. 

\begin{acknowledgments}
We thank the referee for their useful comments and suggestions, which have helped improve the clarity and quality of our manuscript. This work is based on observations collected at the Centro Astron\'omico Hispanoen Andaluc\'ia (CAHA) at Calar Alto, operated jointly by the Andalusian Universities and the Instituto de Astrof\'isica de Andaluc\'ia (CSIC). AP acknowledges funding from the Chinese Academy of Sciences President’s International Fellowship Initiative (PIFI), Grant No. 2024PVC0088. 
This research is supported by the National Key R\&D Program of China (2021YFA1600404 and 2023YFA1607904), by the National Science Foundation of China (NSFC; 11833008, 11991050, 12122305, and 12333003).
\end{acknowledgments}
\vspace{5mm}
\facilities{SDSS, CAHA}

\software{IRAF \citep{1986SPIE..627..733T}}

\appendix
\section{Line flux measurement using simple integration}\label{app:A}
\begin{figure}
    \centering
    \includegraphics[width=10cm, height=16cm, angle =-90]{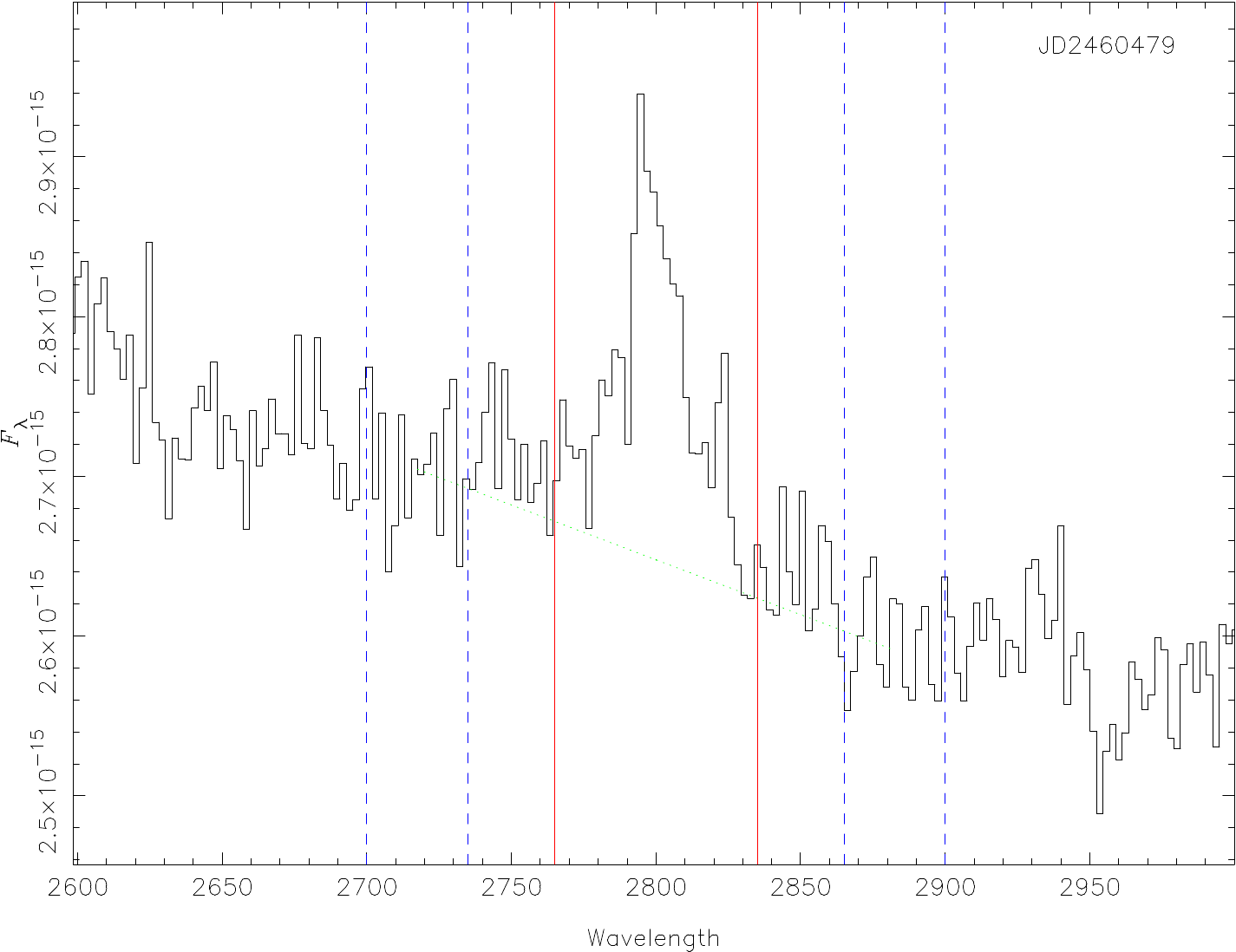}
    \caption{Rest frame spectra of B2 1308+326. Vertical blue dashed lines show the continuum windows while the red solid lines indicate the Mg II line window. }
    \label{fig:mg2}
\end{figure}

We measured the Mg II line flux using simple integration (see Figure \ref{fig:mg2}). First, we established the continuum level by selecting two wavelength windows (shown as blue dashed lines) at 2700–2735 Å and 2865–2900 Å in the rest frame around the Mg II line. Then, we summed the flux above this continuum within the emission line window of 2765–2835 Å (indicated by red solid lines). The estimated integrated flux is (7.3$\pm$0.4) $\times$ 10$^{-15}$ erg cm$^{-2}$ s$^{-1}$.
\bibliography{master}{}
\bibliographystyle{aasjournal}
\end{document}